\documentclass[a4paper,11pt,titlepage]{article}
\usepackage{amssymb}
\newcommand{\eq}[1]{\begin{equation} #1 \end{equation}}
\newcommand{\beq}{\begin{equation}}
\newcommand{\eeq}{\end{equation}}
\newcommand{\bea}{\begin{eqnarray}}
\newcommand{\eea}{\end{eqnarray}}
\newcommand{\rf}[1]{(\ref{#1})}

\newcommand{\nn}{\nonumber}

\begin{document}

\begin{titlepage}

\rightline{NBI-HE-99-16}
\rightline{hep-th/9905067}
\rightline{May, 1999}

\vskip 1cm

\centerline{\Large \bf On $QCD_2$ from supergravity}
\vskip 0.1cm
\centerline{\Large \bf and mass gaps in $QCD$}

\vskip 1cm

\centerline{{\bf Jo\~ao Correia}\footnote{e-mail: correia@nbi.dk} }
\vskip 0.2cm
\centerline{and}
\vskip 0.2cm
\centerline{{\bf Troels Harmark}\footnote{e-mail: harmark@nbi.dk} }
\vskip 0.3cm
\centerline{\sl The Niels Bohr Institute}
\centerline{\sl Blegdamsvej 17, DK-2100 Copenhagen \O, Denmark}

\vskip 2cm

\centerline{\bf Abstract}

\vskip 0.4cm

\noindent
As a test of the conjectured $QCD$/supergravity duality, 
we consider mass gaps in the supergravity construction of $QCD_2$.
We find a mass gap in the dual field theory 
both when using non-rotating and rotating black D2-branes
as backgrounds in the supergravity construction of $QCD_2$.
So, since pure $QCD_2$ does not have a mass gap, the dual field theory
of the supergravity construction of $QCD_2$ cannot be pure $QCD_2$.
Considering the mass scales in the dual field theory 
of the supergravity construction of $QCD_2$,
we find that this is explainable both in the case of the non-rotating 
background and of the rotating background.
In particular, the mass gap in the case of the rotating background 
can be explained using results of the large angular momentum
limit of euclidean rotating branes, obtained recently by Cvetic and Gubser.
We furthermore remark on the possible implications for the mass gaps
in the supergravity constructions of $QCD_3$ and $QCD_4$.

\end{titlepage}


\section{Introduction and summary}

In \cite{maldacena} Maldacena conjectured a duality between large
$N$ superconformal field theories with maximal supersymmetry and
superstring theory or M-theory on certain anti-de Sitter 
backgrounds\footnote{For a recent review of this, see \cite{petersen}.}.
As an extension of this, Witten conjectured in \cite{witten1} 
an approach to study large $N$ non-conformal and 
non-supersymmetric field theories
such as pure $QCD$ (in this paper pure $QCD$ means pure
Yang-Mills theory) at strong coupling
using string theory on certain background geometries.
According to this conjectured duality, 
the large $N$ expansion of the field theory should be identified
to the perturbative expansion of the string theory, and
the strong coupling expansion of the field theory should be identified
to the \(l_s = \sqrt{\alpha'} \) expansion of the string theory.
In particular, one could hope to study $QCD$ for large $N$ 
and for large 't Hooft coupling \( \lambda = g_{YM}^2 N \) via 
a dual supergravity background, being the semiclassical approximation
to weakly coupled string theory.

This idea has received much attention recently 
and it is found that the supergravity description of $QCD$
has many qualitative similarities with the expectations one has of 
pure $QCD$ at strong coupling\footnote{For further discussion see
\cite{petersen} and references therein.}.
However, already from the beginning it has been clear that
to really get to $QCD$, one must take a limit where the supergravity
approximation breaks down, and one must instead consider at least tree-level
string theory in a rather singular background\cite{witten1,gross_ooguri}. 
But, until a better understanding of string theory in these backgrounds 
is reached, one can try to probe how far the supergravity can be used
in understanding field theories at large $N$.

One of the aspects of $QCD$ that has been discussed is the existence
of a mass gap for glueball states. Witten initiated the study of this
question in \cite{witten1}, and it has been studied further 
in several recent papers\cite{massgappapers,hash_oz,ooguri_robins,minahan}.
The upshot of these papers is that $QCD_3$ ($QCD_4$) is constructed
from a background of $N$ black D3-branes (D4-branes) 
in the field theory limit and that the existence of the
mass gap can be shown by considering the dilaton fluctuation equation
in the particular background, since the dilaton couples to the 
$\mbox{tr}(F^2)$ term in the expansion of the Born-Infeld action.

As pointed out in \cite{ooguri_robins,russo}, there are problems in this 
procedure for finding a mass gap in pure $QCD$ from
the supergravity construction of $QCD$. 
The main problem is that the mass gap is of the same order as the
cutoff in the theory. An attempt to solve this was first proposed
in \cite{russo} by introducing rotating black branes, and was later 
extended in \cite{more_russo}. 
The claim of these papers is that for certain limits of 
the angular momentum, the scale of the cutoff should decouple from
the scale of the mass gap. 
This would mean that the fermion mass scale $M_f$ and the mass gap 
scale $M_{gap}$ obey \( M_{gap} << M_f \).
However, as argued in \cite{cvetic_gubser}, this does not seem to hold 
since some of the fermions do not decouple from the mass gap scale
in the considered limit.
So, for both the non-rotating and rotating backgrounds it is not certain
whether a non-zero eigenvalue of the dilaton fluctuation equation 
corresponds to a mass gap in pure $QCD$.

In this paper, we study the supergravity construction of $QCD_2$
for two main purposes:

1) Until now, the proposed $QCD$/supergravity duality 
has almost exclusively been used on 
$QCD_3$ and $QCD_4$, so one of the purposes of this paper is 
to test the proposed duality in the case of pure $QCD_2$(In
\cite{horava} $QCD_2$ from supergravity has been studied, but not the
particular aspects that we consider in this paper).

2) The pure $QCD_2$ theory on a plane has no glueball mass gap, 
since the gluons have no degrees of freedom.
Thus, contrary to the $QCD_3$ and $QCD_4$ cases, 
the supergravity description of $QCD_2$
is not supposed to predict the existence of a mass gap. 
So, in this paper, we will try to test whether 
a mass gap is predicted by the supergravity description.

In this paper, we consider both the non-rotating(see section 2)
and rotating(see section 3) supergravity backgrounds for studying
$QCD_2$. The non-rotating background consists of $N$ black D2-branes in
the field theory limit and the rotating background consists of 
$N$ rotating black D2-branes in the field theory limit.

In this paper, we find that a mass gap is predicted for pure $QCD_2$ for both
non-rotating and rotating backgrounds.
So, for pure $QCD_2$ the supergravity construction of $QCD_2$
has totally different properties than the field theory it was supposed 
to be dual to.
We argue in section 2 and 3 that this is explainable when
considering the mass scales in the field theories dual to the
supergravity backgrounds.
In section 4 we present the conclusions.


\section{Mass gap from non-rotating black D2-branes}

To study $QCD_2$ from supergravity we must construct an appropriate
supergravity background which is dual to a two dimensional non-supersymmetric
field theory. 
To do this, we start by considering $N$ D2-branes in the field theory 
limit\cite{itzhaki_malda}
\[ U \equiv \frac{r}{l_s^2},\ \ l_s \rightarrow 0  \]
The dual field theory on the D2-brane world-volume has 
Yang Mills coupling constant
\[ g_{YM_3}^2 = g_s l_s^{-1} \]
In order to trust the supergravity description,
we must have\cite{itzhaki_malda}
\[ 1 << g_{YM_3}^2 N U^{-1} << N^{4/5} \]
So, we have that \( N >> 1 \) and \( g_{YM_3}^2 N >> U \)
which means that we consider the large $N$ limit of the
$SU(N)$ gauge theory on the world-volume of the D2-branes,
at strong coupling. Since it is the large $N$
limit, we introduce the 't Hooft coupling
\eq{ \lambda_3 \equiv g_{YM_3}^2 N }
%
%
The field theory limit of $N$ black D2-branes gives the 
following solution
\bea
\label{2branethroat}
 \frac{ds^2}{l_s^2} &=& \frac{U^{5/2}}{\sqrt{6\pi^2 \lambda_3}}
\Big[ \Big(1-\frac{U_0^5}{U^5}\Big)dt^2 + (dy^1)^2 + (dy^2)^2 \Big] \nn
\\ && + \frac{\sqrt{6 \pi^2 \lambda_3}}{U^{5/2}} 
\Big(1-\frac{U_0^5}{U^5}\Big)^{-1} dU^2
 + \frac{\sqrt{6 \pi^2 \lambda_3}}{\sqrt{U}} d\Omega_6^2 
\eea
with the dilaton field given by
\eq{ e^\phi = g_{YM_3}^2 (6\pi^2 \lambda_3)^{1/4} U^{-5/4} }
This solution has temperature
\eq{ T = \frac{5}{4\pi}\frac{U_0^{3/2}}{\sqrt{6\pi^2 \lambda_3}} }
One can also start from the 11 dimensional 
supergravity background \( AdS_4 \times S^7 \), with
\( AdS_4 \) being an Anti-de Sitter black hole in 4 dimensions
(this approach to $QCD_2$ was suggested in \cite{horava}).
However, compactifying the \(S^7\) 
on a circle in the usual M-theory/Type IIA
duality manner, one gets the above solution.

The world-volume theory of black D2-branes is a $2+1$ dimensional theory 
described at low energies by the Born-Infeld action. 
It is a non-supersymmetric theory, since the D2-branes are non-extremal and
the fermions acquire masses of order the temperature $T$, 
even before quantum corrections\cite{witten1}.
So, we have a 2 dimensional non-supersymmetric field theory
if we consider energies much smaller than the temperature.
We introduce the two-dimensional Yang-Mills coupling constant
\[ g_{YM_2}^2 = T g_{YM_3}^2 \]
and the 't Hooft coupling
\[ \lambda_2 = T \lambda_3 \]
This is the appropriate coupling constants for the 
dual 2 dimensional field theory.

The coupling of the world-volume fields with the dilaton
can be found by expanding the Born-Infeld action, 
as done in \cite{hash_oz}. From this, one finds that the 
dilaton couples to \( \mbox{tr} (F^2) \).
If we consider the dilaton fluctuation $h$ on the background field $\phi$ 
stated above, we get the dilaton fluctuation equation
\eq{ \label{dilaton}
 \partial_\mu \Big( e^{-2\phi} \sqrt{g} g^{\mu \nu} \partial_\nu h \Big) = 0 }
Introducing the Ansatz
\beq h = f(z) e^{i k_1 y^1 + i k_2 y^2} \label{Ansatz} \eeq
where $z \equiv U / U_0$, equation
(\ref{dilaton}) yields
\beq 
\frac{e^{2\phi}}{\sqrt{g}} g_{y^1 y^1} U_0^{-1} 
\Big( e^{-2\phi} \sqrt{g} g^{UU} U_0^{-1} f' \Big)' 
- k^2 f = 0 
\label{d2} 
\eeq
We identify $k^2 \equiv k_1^2 + k_2^2$ with minus the two-dimensional mass: $m^2=-k^2.$ Reading the metric components from (\ref{2branethroat}) we get
\beq z^{-1} \Big( ( z^6 - z ) f' \Big)' 
+ \frac{25}{16 \pi^2 } \frac{m^2}{T^2} f = 0 
\label{nonrotd}
\eeq 
The dilaton fluctuations are imposed to be regular at $z=1$, and also to be 
square integrable with respect to the metric (\ref{2branethroat}). 
Together, these boundary conditions are sufficient to make a well-posed 
Sturm-Liouville problem which can be solved by standard methods, such as a WKB approximation or a series solution. The result is that $m$ has a discrete spectrum, with a first eigenvalue of 
$m \approx 11 \, T$ (this fact can also be deduced from \cite{minahan}). 

If we proceeded in the same way as in 
\cite{witten1,massgappapers,ooguri_robins,hash_oz,minahan} 
for the $QCD_3$ and $QCD_4$ cases, 
then we would predict a mass gap in pure $QCD_2$.
But, as stated in the introduction, pure $QCD_2$ on a plane has no mass gap
so this procedure cannot be correct.
The reason for this comes from the fact that both the mass gap and
the supersymmetry restoration energy are of order the temperature $T$. 
This means that we cannot expect the fermions to decouple in 
the field theory for which we have computed the mass gap. 
So, it seems that we are not predicting a mass gap in pure $QCD_2$ but
rather in a theory containing fermions.
This also seems to suggest that one cannot conclude anything about 
the existence of a mass gap
in $QCD_3$ and $QCD_4$ from supergravity using non-rotating black D-branes, 
since these theories also predict a mass gap
of order the temperature 
(this was first noted in \cite{ooguri_robins,russo}).


\section{Mass gap from rotating black D2-branes}

An attempt to solve the problem that the mass gap scale
and the fermion mass scale are of the same order
was pointed out by Russo in \cite{russo}, and developed further in
\cite{more_russo}. 
By considering rotating D-branes it was found that one could decouple 
the mass gap scale in $QCD_3$ and $QCD_4$ from the scale of 
the temperature in the limit of very high angular momenta. 

A non-rotating black D2-brane has an $SO(6)$ rotational symmetry 
in the transverse space, so since we can break $SO(6)$ into
$SO(2) \times SO(2) \times SO(2)$, we can have 3 angular 
momentum parameters, $l_1, l_2$ and $l_3$. 
However, as we shall see, we only need one angular momentum \( l=l_1 \)
to decouple the scale of the mass gap from the scale of the temperature.
The rotating black D2-brane solution can be obtained 
from the rotating black hole solutions in an 8 dimensional space-time
as done in \cite{cvetic_duff}. 
To transform this solution with Minkowski signature to a solution 
with euclidean signature, one uses the 
transformation \( t \rightarrow -i\, t \) and \( l \rightarrow i\, l \).
The field theory limit for $N$ rotating D2-branes is\cite{russo,more_russo}
\[ U \equiv \frac{r}{l_s^2},\ \ a \equiv \frac{l}{l_s^2},\ \  
l_s \rightarrow 0 \]
This gives the metric
\bea 
\frac{ds^2}{l_s^2} &=& 
\frac{\sqrt{\Delta} U^{5/2}}{\sqrt{6 \pi^2 \lambda_3}}
\Big[ \Big( 1- \frac{U_0^5}{\Delta U^5}\Big) dt^2 
+ (dy^1)^2 + (dy^2)^2 \Big] \nn \\
&+& \frac{\sqrt{6\pi^2\lambda_3}\sqrt{\Delta}}{ U^{5/2}} 
\Big( 1 - \frac{a^2}{U^2} - \frac{U_0^5}{U^5} \Big)^{-1} dU^2 
+  \frac{\sqrt{6\pi^2\lambda_3}}{\sqrt{\Delta}\sqrt{U}}
\Big[ \Delta d \theta^2 + \cos^2 \theta d \psi_1^2 \nn \\
&+& \cos^2 \theta \cos^2 \psi_1 d \psi_2^2  
+ \Big( 1- \frac{a^2}{U^2}\Big) \sin^2 \theta d \phi_1^2 
+ \cos^2 \theta \sin^2 \psi_1 d \phi_2^2 \nn \\
&+&\cos^2 \theta \cos^2 \psi_1 \sin^2 \psi_1 d \phi_3^2 \Big] 
- 2 \frac{U_0^{5/2}}{\sqrt{\Delta} U^{5/2}} a \sin^2 \theta dt d\phi_1
\label{rot2brane} 
\eea
where
\beq \Delta = 1 - \frac{a^2 \cos^2 \theta}{U^2} \eeq
The dilaton field is given by 
\beq e^\phi =  g_{YM_3}^2 
(6 \pi^2 \lambda_3)^{1/4} \Delta^{-1/4} U^{-5/4} 
\eeq
The temperature can be computed from \rf{rot2brane} to be
\beq 
T = \frac{1}{4 \pi} \frac{1}{\sqrt{6 \pi^2 \lambda_3 }} 
\frac{U_+^4}{U_0^{5/2}} 
\left( 3 + 2 \frac{U_0^5}{U_+^5}\right) \label{trot}
\eeq
where \( U_+ \equiv r_+ / l_s^2 \) and \( r_+ \) is the radius of the outer 
horizon.
The square root of the determinant of the metric \rf{rot2brane} is found to be 
\beq \sqrt{g} = l_s^{10} 6 \pi^2 \lambda_3 U 
\cos^4 \theta  \cos^2 \psi_1  \sin^2 \psi_1 \sin \theta \eeq
It is remarkable that the dilaton fluctuation equation (\ref{dilaton}) 
admits an angle-independent Ansatz of the same form as (\ref{Ansatz}), 
for which it yields 
\beq z^{-1} \Big( z^6 ( 1 - a^2 U_0^{-2} z^{-2}
- z^{-5} ) f' \Big)' 
+  \frac{6 \pi^2 \lambda_3}{U_0^3} m^2 f = 0 
\label{rotd}
\eeq
which reduces to \rf{nonrotd} when $a = 0$ as it should. 
The solutions of this equation are again constrained to be regular 
at the horizon and to be square-integrable with respect to the metric 
\rf{rot2brane}, 
and they can be found by means of a WKB method \cite{more_russo}. 
It turns out that
\beq m^2 = \frac{\pi^2}{\xi^2} k \left( k + \frac{2}{3}\right) \quad k=1,2,... \label{m2} \eeq
where the mass scale is set by
\beq \xi =  \frac{2 \pi}{5} \frac{\sqrt{6 \pi^2 \lambda_3}}{U_0^{3/2}} \int_{z_+}^{\infty} \frac{dz}{\sqrt{z^5 - a^2 U_0^{-2} z^3 - 1}} \label{xi} \eeq
where $z_+ = U_+ / U_0$ is the angular momentum dependent horizon coordinate, given by the largest real solution of
\beq z^5 - a^2 U_0^{-2} z^3- 1 = 0 \label{hor} \eeq

We shall now consider the limit \( a >> U_0 \) to
show that the mass gap scale decouples from the scale of the temperature
in this limit.
For \( a >> U_0 \) we have \( U_+ \sim a \) and 
from \rf{xi} we have \( \xi \sim U_+^{-3/2} \) so we find that the 
$a >> U_0$ limits of \rf{trot} and \rf{m2} are
\beq
T \sim a^4,\ \ 
m \sim a^{3/2}
\eeq
respectively.
This shows that the scale of the mass gap decouples from the scale
of the temperature for \( a >> U_0 \).

Thus, if for a moment we suppose that the scalars and fermions in the D2-brane 
field theory get masses of order the temperature or higher, we are inevitably
lead to the conclusion that pure $QCD_2$ has a mass gap, because
of the decoupling shown above.
Therefore, 
since pure $QCD_2$ does not have a mass gap, the conclusion must be that
some of the fermions or scalars do not decouple from the mass gap scale.
This conclusion has also been reached in \cite{cvetic_gubser}, 
where it was shown 
for rotating M2, M5 and D3-branes that some of the fermions get masses
of order \( \sqrt{1 - a/U_+} \,  T \). 
Since we consider M2-branes with the transverse space compactified 
on a circle, this also applies to our case. 
That some of the fermions are becoming lighter with
higher angular momentum means that a fraction of the original supersymmetry in the non-rotating extremal case is restored for \( a >> U_0 \),
as mentioned in \cite{cvetic_gubser}.
In our case we get 
\beq
 \sqrt{1 - a/U_+} \,  T \sim a^{-5/2} T \sim a^{3/2} 
\eeq
So, some of the fermions have the same mass scale as the mass gap
in this limit.
This means that the mass gap that we have 
obtained for \( a >> U_0 \)
is not a mass gap of pure $QCD_2$, but instead a mass gap in a 
theory containing fermions, just as in the non-rotating case.
As stated in \cite{cvetic_gubser}, 
this result seems to apply also to the pure $QCD_3$
and $QCD_4$ cases, so also in these cases it seems that
one cannot establish the existence
of a mass gap from rotating branes in supergravity.


\section{Conclusions}

In this paper we have considered the correspondence between
$QCD_2$ and supergravity. To test this correspondence, we have 
tried to see if the supergravity construction of $QCD_2$
predicts a mass gap. We found that a mass gap was predicted 
using non-rotating and rotating black D2-branes as supergravity
backgrounds. 
Since pure $QCD_2$ does not have a mass gap, this suggests that
the supergravity construction of $QCD_2$ does not fully reproduce
the dynamics of pure $QCD_2$. 
As argued in sections 2 and 3, the reason for this was that,
for both non-rotating and rotating supergravity backgrounds, 
the dual field theory for which we found a mass gap contained fermions 
and was therefore not pure $QCD_2$.

That we found a mass gap for rotating D2-branes can be seen as
a confirmation of the results of \cite{cvetic_gubser},
since their results precisely explain the discrepancy between
the supergravity construction of $QCD_2$ and pure $QCD_2$.

As already stated, it seems that the problems for the mass gap
calculation in the supergravity construction of $QCD_2$ also
apply for the $QCD_3$ and $QCD_4$ cases, both for non-rotating 
and rotating D-branes. So, as concluded in \cite{cvetic_gubser},
it seems that there is no a priori reason to believe
that the supergravity constructions of $QCD_3$ and $QCD_4$
can say anything about the existence of mass gaps
in pure $QCD_3$ and pure $QCD_4$.


\section*{Acknowledgements}

We would like to thank P. H. Damgaard and J. L. Petersen for many
stimulating discussions and for reading the manuscript.
J.C. acknowledges the support of the EU via grant ERB 4001GT973188.


\end{document}